\documentstyle[12pt]{article} 
\vbox{\vspace{6mm}} 
\renewenvironment{thebibliography}[1]
 { \begin{list}{\arabic{enumi}.}
    {\usecounter{enumi} \setlength{\parsep}{0pt}
     \setlength{\itemsep}{3pt} \settowidth{\labelwidth}{#1.}
     \sloppy
    }}{\end{list}}
\begin{document}                      
\begin{center}
{\large \bf THE RELATION OF CONSTRAINTS ON PARTICLE STATISTICS \\
FOR DIFFERENT SPECIES OF PARTICLES} \\[.1in]

O.W. Greenberg\footnote{Supported in part by a Semester Research Grant from the 
General Research Board of the University of Maryland and by a
grant from the National Science Foundation.}\\
Center for Theoretical Physics\\
Department of Physics\\
University of Maryland\\
College Park, Maryland 20742-4111\\
and\\
Robert C. Hilborn\footnote{On leave from Amherst College}\\
Department of Physics and Astronomy\\
University of Nebraska-Lincoln\\
Lincoln, NE 68588-0111\\
[.4in]
University of Maryland Preprint No. 99-005\\
hep-th/ 
\end{center}
 
To appear in  {\it Foundations of Physics}, special issue in honor of Daniel Mordechai
Greenberger.\\
\vglue 0.8cm
\begin{abstract} 
Quons are particles characterized by the parameter $q$, which permits 
smooth interpolation between Bose and Fermi statistics;
 $q=1$ gives bosons, $q=-1$ gives fermions.
  In this 
paper we give a heuristic argument for an extension of conservation
of statistics to quons with trilinear couplings of the form $\bar{f}fb$, where
$f$ is fermion-like and $b$ is boson-like.  We show that $q_f^2=q_b$.  
In particular, we relate the bound
on $q_{\gamma}$ for photons to the bound on $q_e$ for electrons, allowing the very
precise bound for electrons to be carried over to photons.  An extension of 
this argument suggests
that all particles are fermions or bosons to high precision.
\end{abstract}
\newpage
\vglue 0.6cm
{\bf 1. Introduction}
\vglue 0.2cm
 
It is a great pleasure to contribute a paper to this special issue of {\it
Foundations of Physics} in honor of Daniel Mordechai Greenberger.  Danny's sense of joy
in finding fun in the pursuit of deep understanding of fundamental issues in quantum
mechanics inspires his many friends and colleagues.  We hope this paper will contribute
to the understanding of a basic issue in particle statistics.
 
The connection between spin and statistics is one of the celebrated general results of
quantum mechanics and of quantum field theory.$^{(1-7)}$  Nonetheless, it is fair to point out
that the usual connection, that integer spin particles must be bosons and odd-half-integer
spin particles must be fermions, requires conditions.  In two-dimensional space, there
are anyons.$^{(8,9)}$  A state of two anyons gets an arbitrary phase when the particles are 
transposed, rather than just a plus or minus sign for bosons or fermions.  In higher
dimensions, parabose and parafermi statistics allow many-dimensional representations of 
the symmetric group, rather than the one-dimensional representations that occur for bosons
and fermions.$^{(10,11)}$  Parastatistics is a relativistic theory with observables that 
commute at
spacelike separation.  As such it is a perfectly valid and orthodox theory.  Of course,
it is worth noting that parastatistics is equivalent to a theory with an exact internal 
symmetry.$^{(12-14)}$  The violation of Bose or Fermi statistics in parastatistics is not 
small.
The parabose or parafermi cases of order $p=2$ that are closest to Bose or Fermi 
statistics still allow discrete violations that can be ruled out for a given particle
without doing a precision experiment.
For example, electrons with parafermi statistics of order
two could have two electrons in each quantum state rather than the one of the exclusion
principle.  In that case the entire periodic table of the elements would have very different 
properties.

If one asks ``How well do we know that a given particle obeys Bose or Fermi statistics?," 
we need a quantitative way to answer the question.  That requires a formulation in which either
Bose or Fermi statistics is violated by a small amount.  We cannot just add a small term 
which violates Bose or Fermi statistics to the Hamiltonian; such a term would not be invariant
under permutations of the identical particles and thus would clash with the particles being
identical.  As mentioned above, parastatistics, which does violate Bose or Fermi statistics,
gives gross violations. 

Quons$^{(15,16)}$, labeled by the real parameter $q$, allow a continuous violation of
Bose and Fermi statistics for identical particles, including possible small violations.  
For $q=1$, only the one-dimensional symmetric 
representation of the symmetric group occurs.  For $q=-1$, only the one-dimensional 
antisymmetric 
representation occurs.  For $-1<q<1$ all representations of the symmetric 
group occur.  As $q \rightarrow 1$, the representations with more horizontal (symmetrized) 
boxes in their Young graphs are more heavily weighted; for $q=1$ only the one-dimensional 
symmetric representation survives.  Analogously, as $q \rightarrow -1$, the representations with 
more vertical (antisymmetrized) 
boxes in their Young graphs are more heavily weighted; for $q=-1$ only the one-dimensional 
antisymmetric representation survives.  Thus the departure of $q$ from $1$ for bosons or from
$-1$ for fermions is a measure of the violation of statistics.
Outside the interval $[-1, 1]$, squares of norms of states 
become
negative.  As far as we know, quons are the only case of identical particles in 
three-dimensional
space that has {\it small} violations of Bose and Fermi statistics.

Unfortunately, the quon theory is not completely satisfactory.
The observables in quon theory do not commute at spacelike
separation.  If they did, particle statistics could change continuously from Bose to Fermi
without changing the spin.  Since spacelike commutativity of observables leads to the
spin-statistics theorem, this would be a direct contradiction.  Kinematic Lorentz 
invariance can be maintained, but without spacelike commutativity or anticommutativity
of the fields the theory may not be consistent.  

For nonrelativistic theories, however, quons are consistent.  The nonrelativistic
version of locality is
\begin{equation}
[\rho({\bf x}),\psi({\bf y})]=-\delta({\bf x}-{\bf y})\psi({\bf y})               \label{nrcr}
\end{equation}
for an observable $\rho({\bf x})$ and a field $\psi({\bf y})$ 
and this does hold for quon theories.  It is the antiparticles that prevent locality in
relativistic quon theories.

An earlier article$^{(17)}$ gives a survey of attempts to violate statistics.

\vglue 0.6cm 
{\bf 2. Quons}
\vglue 0.2cm
 
The quon algebra for creation and annihilation operators is
\begin{equation}
a_k a_l ^{\dagger}-q a_l ^{\dagger} a_k=\delta_{kl}.                               \label{qa}
\end{equation}
The parameter $q$ can be different for different particles.
The Fock-like representation that we consider obeys the vacuum condition
\begin{equation}
a_k |0\rangle=0.                                                         \label{vac}
\end{equation}
These two conditions determine all vacuum matrix element of polynomials in the
creation and annihilation operators.  In the case of free quons, all
non-vanishing vacuum matrix elements must have the same number of annihilators
and creators.  For such a matrix element with all annihilators to the left and
creators to the right, the matrix element is a sum of products of 
``contractions'' of the form $\langle 0|a a^{\dagger} |0 \rangle$ just as in the case
of Wick's theorem for
bosons and fermions.  The only difference is that the terms are multiplied by
integer powers of $q$.  The power can be given as a graphical rule:  Put
$\circ$'s
for each annihilator and $\times$'s for each creator in the order in which 
they occur in the matrix element on the x-axis.  
Draw lines above the x-axis connecting the pairs that are
contracted.  The minimum number of times these lines cross 
is the power of $q$ for that term in the matrix element.  

The physical significance of $q$ for small violations of Fermi statistics is
that $q_F=2 v_F-1$, where the parameter $v_F$
that indicates the magnitude of the violation appears in the two-particle density
matrix
\begin{equation}
\rho_2=(1-v_F)\rho_a+v_F \rho_s.
\end{equation}
For small violations of Bose statistics, the two-particle density matrix is
\begin{equation}
\rho_2=(1-v_B) \rho_s+v_B \rho_a,
\end{equation}
where $\rho_{s(a)}$ is the symmetric (antisymmetric) two-boson density matrix.
Then $q_B=1- 2v_B$.

As already stated above, for $q$ in the open interval $(-1, 1)$ all
representations of the symmetric group occur.  As $q \rightarrow 1$, the
symmetric representations are more heavily weighted and at $q=1$ 
only the totally
symmetric representation remains; correspondingly, as $q \rightarrow -1$, the
antisymmetric representations are more heavily weighted and at $q=-1$ only the 
totally antisymmetric representation remains.  Thus for a general 
$n$-quon state, there
are $n!$ linearly independent states for $-1<q<1$, but there is only one 
state for $q= \pm 1$.

We emphasize something that some people find very strange: {\it there is no
operator
commutation relation between two creation or between two annihilation 
operators,} except for $q= \pm 1$, which, of course, correspond to Bose and 
Fermi statistics.  Indeed, the fact that the general $n$-particle state with
different quantum numbers for all the quons has $n!$ linearly independent
states proves that there is no such commutation relation between any number 
of creation
(or annihilation) operators.  (An even stronger statement holds:  There is no 
two-sided ideal containing a term with only creation or only annihilation operators.)
 
Quons are an operator realization of ``infinite statistics'' that was found as
a possible statistics by Doplicher, Haag and Roberts$^{(13)}$ in their general
classification of particle statistics.  The simplest case, $q=0$,$^{(15)}$,
suggested to one of the authors (OWG) by Hegstrom,$^{(18)}$ 
was discussed earlier in the context of operator algebras by Cuntz.$^{(19)}$
It seems
likely that the Fock-like representations of quons for $|q|<1$ are homotopic to
each other and, in particular, to the $q=0$ case, which is particularly simple.
All
bilinear observables can be constructed from the number operator, 
$n_k$, defined by
\begin{equation}
[n_k, a^{\dagger}_l]_-=\delta_{kl}a^{\dagger}_l,                                  \label{num}
\end{equation}
or the transition operator, $n_{kl}$, defined by
\begin{equation}
[n_{kl}, a^{\dagger}_m]_-=\delta_{lm}a^{\dagger}_k.                              \label{trans}
\end{equation}
Clearly, $n_k \equiv n_{kk}$. 
For $q \neq \pm 1$, these operators are represented by infinite series in the creation and
annihilation operators.
Once Eq.(6 or 7) holds, the Hamiltonian and other observables can be
constructed in the usual way; for example,
\begin{equation}
H=\sum_k \epsilon_k n_k.                                                 \label{11}
\end{equation}

As in the Bose case, the transition or number 
operator for quons defines an unbounded operator whose domain includes states made by
polynomials in the creation operators acting on the vacuum.  What is different for quons is that
the number operator has an infinite degree representation in terms of the annihilation and 
creation operators.
(As far as we know, this is the first case in which the number operator, 
Hamiltonian, etc. for a free field are of infinite degree.  Presumably this is
due to the fact that quons are a deformation of an algebra and are related to
quantum groups.)  Several authors$^{(20-23)}$ gave
proofs of the positivity of the squares of norms of quon states.
It is amusing to note that, despite the lack of locality or antilocality, 
the free quon field obeys the TCP
theorem and Wick's theorem (with added factors of $q$ as mentioned above) holds for quon 
fields.$^{(16)}$

\vglue 0.6cm 
{\bf 3. The Ramberg-Snow bound on $v_F$ for electrons}
\vglue 0.2cm

Ramberg and Snow$^{(24)}$ were able to place an extremely
high-precision bound on possible violations of Fermi statistics for electrons by passing
a 30 ampere current through a thin strip of copper for a month and looking for anomalous
x-rays due to transitions of conduction electrons falling into the K-shell of the copper
atoms, which could occur if the conduction electrons violated the exclusion principle
by a small amount and thus were not always antisymmetric with respect
to the electrons in the copper atoms.  Avogadro's number is on their side.  If the 
spin-statistics connection is violated, a given collection of
electrons can, with different probabilities, be in different permutation
symmetry states.  The probability to be in the ``normal'' totally antisymmetric
state would presumably be close to one, 
the next largest probability would occur for the
state with its Young tableau having one row with two boxes, etc.  The idea of
the experiment is that each collection of electrons has a possibility of being
in an ``abnormal'' permutation state.  If the density matrix for a conduction 
electron together with the electrons in an atom has a projection onto such an
``abnormal'' state, then the conduction electron will not see the K shell of
that atom as filled. A transition into the K shell with x-ray emission
is then allowed.  Each conduction electron that comes sufficiently close to a given
atom has an independent chance to make such an x-ray-emitting transition, and
thus the probability of seeing such an x-ray is proportional to the number of
conduction electrons that traverse the sample and the number of atoms that the
electrons visit, as well as the probability that a collection of electrons can
be in the anomalous state.   Ramberg and Snow estimated the energy of the
x-rays that would be emitted due to the transition to the K shell and found no
excess of x-rays above background in this energy region.  They set the limit 
\begin{equation}
v_e \leq 1.7 \times 10^{-26}.                  
\end{equation}
This is high precision, indeed!
 
This experiment succeeded because systems
of several electrons occur in bound states in atoms and because individual x-ray transitions
could be detected.  The corresponding experiment for
photons is impossible, because photons obey Bose statistics so that adding an additional
photon to a state of several photons will not produce a sharp signal, such as an x-ray.
Also, photons do not occur in bound states such as electrons in atoms. 

\vglue 0.6cm 
{\bf 4. Conservation of statistics}
\vglue 0.2cm

The first conservation of statistics theorem states that terms in the Hamiltonian density 
must have an even number of Fermi fields and that composites of fermions and bosons are 
bosons, unless they contain an odd number of fermions, in which case they are 
fermions.$^{(25,26)}$  
The extension to parabosons and parafermions is more complicated;$^{(10)}$ however, the main
constraint is that for each order $p$ at least two para particles must enter into every
reaction. 

Reference (27) argues that the condition that the energy of widely
separated subsystems be additive requires that all terms in the
Hamiltonian be ``effective Bose operators'' in that sense that
\begin{equation}
[{\cal H}({\bf x}), \phi({\bf y})]_-\rightarrow 0, |{\bf x}-{\bf y}| \rightarrow \infty.                                          \label{ascr}
\end{equation}
For example,  ${\cal H}$ should not have a term such as 
$\phi(x){\psi(x)}$, where ${\phi}$ is 
Bose and $\psi$ is Fermi, because then the contributions to the 
energy of widely separated subsystems would alternate in sign.  Such terms are
also prohibited by rotational symmetry.  This discussion was given in the context of
external sources.  For a fully quantized field theory, one can replace Eq.(10)
by the asymptotic causality condition, asymptotic local commutativity,
\begin{equation}
[{\cal H}({\bf x}), {\cal H}({\bf y})]_-=0, |{\bf x}-{\bf y}| \rightarrow \infty \label{decr}
\end{equation}
or by the stronger causality condition, local commutativity,
\begin{equation}
[{\cal H}({\bf x}), {\cal H}({\bf y})]_-=0, {\bf x} \neq {\bf y}.           \label{decr2} 
\end{equation}
Studying this condition for quons in electrodynamics is complicated, 
since the terms in the interaction
density will be cubic.  It is simpler to use the description of the electron current or
transition operator as an external source represented by a quonic Grassmann number.
  
Here we 
give a heuristic argument for conservation of statistics for quons based on a simpler
requirement in the context of quonic Grassmann external sources.$^{(27)}$
The commutation relation of the quonic photon operator is          
\begin{equation}
a(k) a(l) ^{\dagger}-q_{\gamma} a(l) ^{\dagger} a(k)=\delta(k-l),
\end{equation}
where $q_{\gamma}$ is the q-parameter for the photon quon field.
We call the 
quonic Grassmann numbers for the electron transitions to which the photon quon operators
couple $c(k)$.
The Grassmann numbers that serve
as the external source for coupling
to the quon field for the photon must obey
\begin{equation}
c(k) c(l)^{\star}  -q_{\gamma} c(l)^{\star}  c(k)=0,       \label{grass}
\end{equation}
and the relative commutation relations must be
\begin{equation}
a(k) c(l)^{\star} -q_{\gamma} c(l)^{\star}  a(k)=0,        \label{relative}
\end{equation}
etc.  Since the electron current
for emission or absorption of a photon with transition of the electron from one atomic
state to another is bilinear in the creation and annihilation operators for the electron,
a more detailed description of the photon emission would treat the photon as
coupled to the electron current, rather than to an external source.  
We impose the requirement that the leading terms in the commutation relation for the
quonic Grassmann numbers of the source that couples to the photon should be mimicked by 
terms bilinear in the electron operators. The electron operators obey the relation
\begin{equation}
b(k) b^{\dagger}(l)- q_{e}b^{\dagger}(l) b(k)=\delta(k-l),     \label{qe}
\end{equation}
where $q_e$ is the q-parameter for the electron quon field.  

To find the connection between $q_e$ and $q_{\gamma}$ we make the following associations,
\begin{equation}
c(k) \Rightarrow b^{\dagger}(p)b(k+p), ~~ c^{\star}(l) \Rightarrow b^{\dagger}(l+r)b(r)
                                                                   \label{replace}
\end{equation}
We now replace the $c$'s in Eq.(14) with the products of operators given in 
Eq.(17) and obtain
\begin{equation}
[b^{\dagger}(p) b(k+p)] [b^{\dagger}(l+r) b(r)]
- q_{\gamma} [b^{\dagger}(l+r) b(r)][b^{\dagger}(p) b(k+p)]=0.      \label{ newcr1}                                  
\end{equation}
This means that the source $c(k)$
is replaced by a product of $b$'s that destroys net momentum $k$; the source $c^{\star}(l)$
is replaced by a product of $b$'s that creates net momentum $l$.  
We want to rearrange the operators in the first term of Eq.(18) to match the second
term, because this corresponds to the standard normal ordering for the transition operators.
For the products $b^{\dagger} b$ we use Eq.(16).  For the products $bb$, as mentioned
above, there is no operator relation; however {\it on states in the Fock-like representation}
there is an approximate relation,
\begin{equation}
b(k+p) b(r) = q_e b(r) b(k+p) + {\rm ~terms ~of ~order} ~~1-q_e^2.         \label{approx}
\end{equation}
In other words, in the limit $q_e \rightarrow -1$, we retrieve the usual anticommutators        
for the electron operators.  (The analogous relation for an operator that is approximately
bosonic would be that the operators commute in the limit $q_{bosonic} \rightarrow 1$.)
We also use the adjoint relation
\begin{equation}
b^{\dagger}(p) b^{\dagger}(l+r)=q_eb^{\dagger}(l+r) b^{\dagger}(p) + 
{\rm terms ~of ~order} ~~1-q_e^2
\end{equation}
and, finally,
\begin{equation} 
q_e b^{\dagger}(p) b(r)=b(r) b^{\dagger}(p) + {\rm terms ~of ~order} ~~1-q_e^2.
\end{equation}
We require only that the quartic terms that correspond to the quonic Grassmann relation
Eq.(14) cancel, so we drop terms in which either $k+p=l+r$ or $r=p$. We also drop
terms of order $1-q_e^2$.  In this approximation, we 
find that Eq.(18) becomes
\begin{equation}
(q_e^2-q_{\gamma})[b^{\dagger}(l+r) b(r)][b^{\dagger}(p) b(k+p)] \approx 0,  \label{newapprox}                             \label{newcr}
\end{equation}
and conclude that 
\begin{equation}
q_e^2 \approx q_{\gamma}.           \label{result}
\end{equation}
This relates the bound on violations of Fermi statistics for electrons to the bound on violations
of Bose statistics for photons and allows the extremely precise bound on
possible violations of Fermi statistics for 
electrons to be carried over to photons.  Eq.(23)
is the quon analog of the conservation of statistics relation that the square of the
phase for transposition of a pair of fermions equals the phase for 
transposition of a pair of bosons.

Arguments analogous to those just given, based on the source-quonic photon relation, 
Eq.(15), lead to 
\begin{equation}
q_{e\gamma}^2 \approx q_{\gamma},
\end{equation}
where $q_{e\gamma}$ occurs in the relative commutation relation
\begin{equation}
a(k) b^{\dagger}(l)=q_{e\gamma}^2 b^{\dagger}(l) a(k).
\end{equation}
Since the normal commutation relation between Bose and Fermi fields is 
for them to commute$^{(28)}$,
this shows that $q_{e\gamma}$ is close to one.

Since the Ramberg-Snow bound on Fermi statistics for electrons is 
\begin{equation}
v_e \leq 1.7 \times 10^{-26}  \Longleftrightarrow  q_e \leq -1 + 3.4 \times 10^{-26},
\end{equation}  
the bound on Bose statistics for photons is
\begin{equation}
q_{\gamma} \geq 1 - 6.8 \times 10^{-26}  \Longleftrightarrow 
v_{\gamma} \leq 3.4 \times 10^{-26}.
\end{equation} 
This
bound for photons is much stronger than could be gotten by a direct experiment.
Nonetheless, 
D. Budker and D. DeMille are performing an experiment that promises to give the best
{\it direct} bound on Bose statistics for photons.$^{(29)}$. 
It is essential to test every basic property in as direct a way as possible.  Thus 
experiments that yield direct
bounds on photon statistics, such as the one being carried out by Budker and 
DeMille, are important.

Teplitz, Mohapatra and Baron have suggested a method to set a very low limit on
violation of the Pauli exclusion principle for neutrons.$^{(30)}$.

The argument just given that the $q_e$ value for electrons implies $q_{\gamma}\approx
q_e^2$ for photons
can be run in the opposite direction to find 
$q{_\phi}^2 \approx q_{\gamma}$ for each charged field
$\phi$ that couples bilinearly to photons.  Isospin and other symmetry arguments then imply 
that
almost all particles obey Bose or Fermi statistics to a precision comparable to the 
precision with
which electrons obey Fermi statistics.

In concluding, we note that further work should be carried out to justify the approximations
made in deriving Eq.(24) and also
to derive the relations among the $q$-parameters that follow from couplings that do not 
have the form $\bar{f}fb$.  We plan to return to this topic in a later paper.

\vglue .6cm
{\bf Acknowledgements}
\vglue 0.2cm

OWG thanks the Aspen Center for Physics for a visit during which part of
this work was carried out.  The unique atmosphere of the Center encourages concentrated
work, without the distractions of one's home university. This work was supported in part 
by the National Science Foundation.  RCH thanks the University of Nebraska-Lincoln for
hospitality and support and Amherst College for support during a sabbatical leave during
which this work was done.  We thank Joseph Sucher for a careful reading of a
draft of this paper and for several helpful suggestions.

\vspace{.6cm}
{\bf References}
\vglue .2cm

\end{document}